\documentstyle[times,pramana,epsf,floats]{ias}

\def\PRD#1{{ Phys.\ Rev.} {\bf D#1}}

\def\vev#1{\langle #1 \rangle}
\def\sss{\scriptscriptstyle}
\def\Ls{{\sss L}}
\def\Ms{{\sss M}}
\def\Rs{{\sss R}}
\def\Tr{\rm Tr}
\newcommand{\nc}{\newcommand}
\nc{\beq}{\begin{equation}}
\nc{\eeq}{\end{equation}}

\newcommand{\be}{\begin{equation}}
\newcommand{\ee}{\end{equation}}
\newcommand{\bea}{\begin{eqnarray}}
\newcommand{\eea}{\end{eqnarray}}
\newcommand{\barr}{\begin{array}}
\newcommand{\earr}{\end{array}}

\begin{document}
\mark{{Neutrino and astroparticle physics ...}{S Mohanty  and U A Yajnik}}
\title{Neutrino and astroparticle physics : Working group report }

\author{S.  MOHANTY and U. A. {$\rm YAJNIK^*$} }
\address{Physical Research Laboratory, Navarangpura, Ahmedabad
380\thinspace009, India \\ 
${}^*$Indian Institute of Technology, Powai, Mumbai 400\thinspace076,
India}

\abstract{The contributions made to the Working Group activities on
neutrino and astroparticle physics are summarised in this article.
The topics discussed were inflationary models in Raman-Sundrum
scenarios, ultra high energy cosmic rays and neutrino oscillations in 4
flavour and decaying neutrino models.}

\maketitle

\section{Leptogenesis in the Left-Right Symmetric Model}
\begin{center}
J. Cline and U. A. Yajnik
\end{center}

It was shown in \cite{lrwhp5} and \cite{ywmmc} that topological defects
such as domain walls and  cosmic strings are generic to the left-right
(L-R) symmetric model with two triplet and one bidoublet Higgs fields. In
particular, the discrete L-R symmetry of the triplet Higgs potential
implies that at the  scale $v_\Rs$, in a given region in the Universe,
either the $SU(2)_L$ or the $SU(2)_R$ could break. Immediately after
this phase transition therefore we would find domains with either of
the above groups broken. Such domains will be separated by walls,
dubbed L-R walls in \cite{ywmmc}.  Phenomenology demands that the
$SU(2)_R$ bosons must eventually acquire the larger masses, and that
the domain structure disappears in order to not dominate the energy
density in the Universe.  This will be achieved if GUT scale physics
induces small corrections to the triplet Higgs potential, making it
energetically favourable for the $SU(2)_L$ bosons to remain the lighter
species. This will be assumed in the following.

Recently with the increasing difficulties in explaining the baryon
asymmetry either in the Standard Model or its minimal supersymmetric
extension, the idea of leptogensis which then produces
the baryon asymmetry has gained ground. In the L-R model,
the disappearance of the domains with broken $SU(2)_L$ provides
a preferred direction for the motion of the domain walls.  This
can fulfill the out-of-thermal-equilibrium requirement needed for
leptogenesis.

Consider the interaction of neutrinos from the L-R wall which is moving
into the energetically disfavored phase.  The left-handed neutrinos,
$\nu_\Ls$,  are massive in this domain, whereas they are massless in
the phase behind the wall. This can be seen from the Majorana mass term
$h_\Ms\Delta_\Ls{\overline{\nu_\Ls^c}}\nu_\Ls$, and the fact that
$\vev{\Delta_\Ls}$ has a kink-like profile, being zero behind the wall
and $O(v_\Rs)$ in front of it.

To get leptogenesis, one needs an asymmetry in the reflection and
transmission coefficients from the wall between $\nu_\Ls$ and its CP
conjugate $(\nu_\Ls^c)^*$. This can happen if a CP-violating condensate
exists in the wall as discussed below. Then there will be a preferance
for transmission of, say, $\nu_\Ls$. The corresponding excess of
antineutrinos $(\nu_\Ls^c)$ reflected in front of the wall will quickly
eqilibrate with $\nu_\Ls$ due to helicity-flipping scatterings, whose
amplitude is proportional to the large Majorana mass.  However the
transmitted excess of $\nu_\Ls$ survives because it is not coupled to
its CP conjugate in the region behind the wall, where $\vev
{\Delta_L}=0$.

A quantitative analysis of this effect can be made either in the
framework of quantum mechanical reflection, valid for domain walls
which are narrow compared to the particles' thermal de Broglie
wavelengths, or using the classical force method \cite{jpt}, which is
appropriate for walls with larger widths.  We adopt the latter here.
The classical CP-violating force of the wall on the neutrinos, 
whose sign is different for $\nu_\Ls$ and $\nu_\Ls^c$, is
\beq
\label{eq:force}
F = \pm \frac{1}{2E^2}\left(m_\nu^2(x) \alpha'(x)\right)'
\eeq
where $m_\nu^2(x)$ is the position-dependent mass and $\alpha$ is
the spatially varying CP-violating phase. One can then write
a diffusion equation for the chemical potential $\mu$ of the 
$\nu_\Ls$ as seen in the wall rest frame
\beq
\label{eq:diffeq}
-D_\nu \mu'' - v_w \mu'
+ \theta(x)\, \Gamma_{\rm hf}\,\mu = S(x)
\eeq
Here $D_\nu$ is the neutrino diffusion coefficient,
$v_w$ is the wall velocity, $\Gamma_{\rm hf}$ is the rate of helicity
flipping interactions taking place in front of the wall (hence
the step function $\theta(x)$), and $S$ is the source term,
given by
\beq
\label{eq:source}
   S(x) = - {v_w D_\nu \over \vev{\vec v^{\,2}}} 
	\vev{v_x F(x)}',
\eeq
where $\vec v$ is the neutrino velocity and 
the angular brackets indicate thermal averages. 

The spatially varying complex neutrino mass can be found from
the finite-temperature effective potential for the L-R model,
thus specifying the source (\ref{eq:force},\ref{eq:source}).
The diffusion equation (\ref{eq:diffeq}) can be solved using
standard Green's function methods \cite{ck}.

\subsection{CP violation}
An attractive feature of the Left-Right symmetric model is
dynamical generation of the CP violation \cite{barenber}. 
After accounting for the phases that can be eliminated by
global symmetries and field redefinitions, the two remaining
phases can be introduced into the VEVs \cite{barenber,dgko}
$$
\vev{\Delta_\Rs} = {1\over\sqrt2}\pmatrix{0 &0\cr v_\Rs e^{i\theta} &0\cr}
\quad{\rm and }\quad 
{\vev\phi} = {1\over\sqrt2}\pmatrix{k_1e^{i\alpha} &0\cr 0 &k_2\cr}
$$
Here $\phi$ is the bidoublet and we use $\tilde\phi$ for its
$SU(2)$ conjugate matrix.
We need the effective CP violating phase to enter as a position
dependent function in the wall profile. It can be shown that the
terms 
$$
\beta_1 \Tr(\phi\Delta_\Rs\phi^{\dagger}\Delta_\Ls^{\dagger})
+ \beta_2 \Tr({\tilde\phi}\Delta_\Rs\phi{\dagger}\Delta_\Ls^{\dagger})
+ \beta_3 \Tr(\phi\Delta_\Rs{\tilde\phi}^{\dagger}\Delta_\Ls^{\dagger})
+{\rm h.c.}
$$
give rise to a potential for the phases $\alpha$ and $\theta$ of 
the form
$$
\beta_1 k_1k_2v_\Ls v_\Rs \cos(\alpha-\theta)
+\beta_2  k_1^2v_\Ls v_\Rs \cos(2\alpha-\theta)
+\beta_3   k_2^2v_\Ls v_\Rs \cos\theta
$$
The bidoublet is not expected to acquire a VEV at the temperature 
scale of $v_\Rs$. However, in the presence of the position dependent
VEV of the triplets, it can turn on in the wall interior. Analysis
of the above potential shows that this gives rise to position
dependent VEV's for the two phases as well. 

Thus our preliminary analysis gives a strong indication of feasible
leptogenesis during the epoch of disappearance of the L-R walls, which
are generic to the Left-Right symmetric model. Depending on the
parameters of the model, the lepton asymmetry could be quite large.
If it is distributed in a certain way between the different lepton
flavors, this large lepton asymmetry can be converted by sphalerons
into a small baryon asymmetry \cite{mrmr}.  Otherwise one expects
a baryon asymmetry of the same order of magnitude as the lepton
asymmetry.

\section{\bf Inflation with bulk fields in the Randall-Sundrum warped
compactification?}
\begin{center}
J. Cline and U. A. Yajnik
\end{center}

\subsection{Introduction}
\label{sec:intro}

The Randall-Sundrum proposal for solving the hierarchy problem has
received much attention in the last year \cite{RS}.  
They considered an extra
compact dimension with coordinate $y$ and line element
\beq
\label{eq:ds2}
	ds^2 = a^2(y) \eta_{\mu\nu} dx^\mu dx^\nu + b^2 dy^2
\eeq
where $a(y) = e^{-kb|y|}$, $y\in[-1,1]$ and the points $y$ and $-y$ are
identified so the extra dimension is an orbifold with fixed points at
$y=0$ and $y=1$.  The four-dimensional geometry is conformal to
Minkowski space.  The scale $k$ in the warp factor $a(y)$ is determined
by the 5-D cosmological constant, $\Lambda$ and the analog to the
Planck mass, $M$, by $k = (-\Lambda/6M^3)^{1/2}$.  Hence $\Lambda$ must
be negative, and the 5-D space is anti-deSitter.

At $y=0$ there is a positive tension brane (the Planck brane) on which
particle masses are naturally of order the Planck mass, $M_p$, while at
$y=1$ there is a negative tension brane (called the TeV brane), where
particle masses are suppressed by the warp factor $e^{-kb}$.  By
adjusting the size of the extra dimension, $b$, so that $kb\sim 37$, the
masses of particles on the TeV brane will be in the TeV range, even if
all the underlying mass parameters (including $\Lambda$, $M$ and $k$)
are of order $M_p$ to the appropriate power.

Although this is desirable for solving the hierarchy problem, it makes
it difficult to understand the origin of inflation.  The density
perturbations from inflation, $\delta\rho/\rho$, are suppressed by
inverse powers of $M_p$.  To get $\delta\rho/\rho\sim 10^{-5}$, one
needs a mass scale much larger than 1 TeV in the numerator.  By
construction, the TeV scale is the cutoff on the TeV brane, so it is hard
to see where such a scale could come from unless some physics outside of
the TeV brane is invoked.

\subsection{Inflation with a bulk scalar}

In this study we investigate what happens when the inflaton is a bulk
scalar field.  The simplest possibility is chaotic inflation with a
free field \cite{linde1}.  We will assume that the line element
(\ref{eq:ds2}) is modified by replacing the Minkowski metric with 4-D
deSitter space, whose line element is $ds^2 = -dt^2 + e^{2Ht} d\vec
x^2$.  The action for the bulk scalar is then
\beq
	S = \frac12 \int d^{\,4}x\, dy\, b\, e^{3Ht} a^4(y) \left[ a^{-2}(y)
\dot\phi^2
	- b^{-2} \phi'^2 - m^2\phi^2\right]
\eeq
The equation of motion for $\phi$ is
\beq
	a^{-2}\left(\ddot\phi + 3H\dot\phi\right) - b^{-2}
	\left(\phi''-4kb\phi'\right)+m^2\phi = 0
\eeq
and, assuming that the size of the extra dimension is stabilized
\cite{CGRT}, the Hubble rate is given by
\beq
	H^2 = {4\pi G\over 3} \int_0^1 dy\,b\,a^4(y)\left(
	a^{-2}\dot\phi^2 + b^{-2}\phi'^2 + m^2\phi^2\right),
\eeq
where $G$ is the ordinary 4-D Newton's constant.

We look for a separable solution, $\phi = \phi_0(t) f(y)$.  We take
$\phi_0$ to have dimensions of (mass)$^1$ so that it is the canonically
normalized field in an effective 4-D description, and $f$ has dimensions of
(mass)$^{1/2}$.  We will also assume the slow roll condition is fulfilled
so that the terms $\ddot\phi$ and $\dot\phi^2$ can be ignored in the last
two equations.  The equation of motion becomes
\beq
	\dot\phi_0 = {e^{-2kby}\over 3\hat H}\left(
	-m^2 + {1\over b^2 f}\left(f''-4kbf'\right)\right) = 
	{\rm constant} \equiv -\Omega
\eeq
with
\beq
\label{eq:HH}
	\hat H^2 \equiv {H^2\over\phi_0^2} = 
	{4\pi G\over 3} \int_0^1 dy\,b\,e^{-4kby}\left(
         b^{-2}f'^2 + m^2 f^2\right).
\eeq
The solution for $\phi_0$ is obviously linear, $\phi_0(t) = C - \Omega
t$.  For chaotic inflation we want $C \gg M_p$ (necessary to fulfill
the slow roll condition \cite{linde1}) and $\Omega>0$, so that $\phi_0$
is rolling to the minimum of its potential.

The equation for $f$ becomes
\beq
\label{eq:eom}
	f'' - 4kb f' - b^2\left(m^2 - 3\hat H\Omega e^{2kby}
	\right)f = 0,
\eeq
This is the same equation as (7) of ref.\ \cite{GW1}.  The solutions
$f_n$ (called $y_n$ in \cite{GW1}) are discrete, such that $\hat H\Omega$
is quantized:
\beq
\label{eq:ev}
	3 k^{-2}\hat H\Omega e^{2kb} = x_{n\nu}^2
\eeq
Here $x_{n\nu}$ is the $n$th root of the equation $2J_\nu(x_{n\nu})
+ x_{n\nu}J'_\nu(x_{n\nu}) = 0$, where the order of the Bessel function
is $\nu=\sqrt{4+m^2/k^2}$.  For $m/k$ in the range $0.5-3$, the lowest
mode $x_{1,\nu}$ ranges from 4 to 6.  This assumes the boundary condition that
$f'=0$ at $y=0,1$, but other choices of boundary conditions will lead to
essentially identical conclusions, as we will explain below.  The $f_n$'s
are normalized so that 
\beq
\label{eq:norm}
	\int_0^1 dy\,b\,e^{-2kby} f_n(y) f_m(y) = \delta_{mn}
\eeq

With the solution for $f_n$ we can evaluate the rescaled Hubble rate,
$\hat H$, in eq.\ (\ref{eq:HH}).  After a partial integration and use
of the equation of motion (\ref{eq:eom}), the integral in (\ref{eq:HH})
becomes identical to that of (\ref{eq:norm}), times $3\hat H\Omega$.
This gives $\hat H^2 = 4\pi G \hat H \Omega$, which together with
eq.\ (\ref{eq:ev}) determines $\Omega$, the rate at which $\phi_0$
is rolling to its minimum:
\beq
	\Omega = {k x_{n\nu} e^{-kb}\over \sqrt{12\pi G}}
	\sim M_p \times {\rm 1 TeV}
\eeq
We used the fact that $e^{-kb}$ is supposed to be of order (TeV)$/M_p$.

\subsection{Density perturbations}

We can now estimate the magnitude of density perturbations in this model.
Using $\delta\rho/\rho\sim H^2/|\dot\phi_0|$, 
\beq
	{\delta\rho\over\rho}\sim (4\pi G)^2\, \Omega\, (C-\Omega t)^2
	\sim {{\rm TeV}\over M_p}
\eeq
Although $C$ is presumed to be super-Planckian, $(C-\Omega t)$ will not be
orders of magnitude larger than $M_p$ near the end of inflation, when the
perturbations with COBE-scale wavelengths were being produced; hence we
take $(C-\Omega t)\sim M_p$ in the above estimate.   This suppression 
of the density perturbations makes our model not viable.

Nevertheless, it is interesting to compare to what would happen if we
tried to do chaotic inflation using a scalar field trapped on the TeV
brane.  The mass of the field is now constrained to be of order $m
\sim$ TeV because of the suppression of masses by the warp factor.
The equation of motion during the slow roll regime is
\beq
	\dot\phi = -{m^2\phi\over 3H} = - {m^2\phi\over \sqrt{
	4\pi G m^2\phi^2/3} };
\eeq
hence $\phi$ evolves linearly with time, and $\dot\phi$ is of order 
$M_p\times 1 $TeV, just as with the bulk scalar field.  And the estimate
for $\delta\rho/\rho$ has the same parametric form.  All this, despite the
fact that we started with a bulk scalar whose mass is Planck-scale in the
5-D Lagrangian.

In retrospect, this result is not surprising.  Reference \cite{GW1}
noted that the modes of the bulk scalar behave similarly to TeV-scale
particles on the brane.  This can be understood by the form of the
solutions \cite{GW1},
\beq
	f_n \sim e^{2ky} J_\nu\left(x_{n\nu}
	e^{kb(y-1)}\right),
\eeq
which are strongly peaked near $y=1$.  There is thus little practical
difference between the low-lying modes of the bulk field and a field
confined to the TeV brane.  

One might wonder if this conclusion depends on the choice of boundary
conditions for the modes $f_n$.  However the fact that the modes peak
at the TeV brane comes from bulk energetics, not boundary conditions.
Since the mass of the bulk field is effectively varying like $e^{-kby}$,
it is energetically much more efficient for the field to be concentrated
near $y=1$.

\subsection{Conclusion}

The simplest chaotic inflation models seem to be ruled out in the
Randall-Sundrum scenario, whether the inflaton is a bulk field or one
restricted to the TeV brane.  One could, alternatively, put the
inflaton on the Planck brane (at $y=0$), at the cost of reintroducing a
hierarchy problem---why should $m/M_P$ be $O(\delta\rho/\rho)\sim
10^{-5}$?  This fine-tuning problem always occurs in inflation, but the
RS setting casts it in a somewhat new light.  One could invent an
intermediate brane for the inflaton, which has just the right mass
scale, but this seems artificial.  Perhaps the RS idea, if correct, is
telling us that hybrid inflation (involving more than one field) is
necessary.

\section{A parametrization for vacuum mixing matrix in four generations}
\begin{center}
{Sandhya Choubey, Gautam Dutta, Srubabati Goswami,
D. Indumathi, 
Debashis Majumdar, M. V. N. Murthy}
\end{center}

Many participants mentioned above, either together or separately, are
studying the solar, atmospheric and supernova neutrinos and their
signatures at the earthborne detectors. It has become clear, since the
announcement of the LSND results, that one may need a fourth sterile
neutrino in order to account for all the experimental data. During the
discussions the participants attempted to arrive at a parametrization for
the vacuum mixing matrix which would be used by all the members in
analysing the data from different detectors as well as different phenomena
involving neutrinos. This would then facilitate comparison of the results
emanating from different groups.  Basically the following constraints
arising from experiments were imposed. 

\begin{enumerate}
\item The solar and atmospheric neutrino data indicate the existence of
two very different scales, $10^{-6}~eV^2$ and $10^{-3}~eV^2$, for the mass 
squared differences between neutrinos. However LSND data requires in 
addition a scale in the range of $eV^2$, different from both atmospheric 
and solar neutrino results. The favoured scenario is that in which there 
are two neutrino doublets. The mass squared difference in each doublet is 
given by the solar and atmospheric neutrino requirements. The doublets 
themselves are separated by a scale corresponding to the LSND data.  

\item While this defines the structure of the mixing matrix, further 
constraint is imposed using the results from the {\sc chooz} reactor
neutrino 
experiment. This result basically imposes an upper limit for the 
conversion of electron anti-neutrino to any other flavour. The limit on 
the relevant mixing parameter is $\epsilon \le 0.07$,

\end{enumerate}

We make use of the fact that the parameter $\epsilon$ is small compared to
unity, and parametrize the unitary mixing matrix. The mixing matrix, which
relates the flavour and mass eigenstates in the four generation scenario
has six angles and the CP-violating phases which we do not consider here.
As is the convention, we denote the mixing angle in the lower doublet by
$\omega$ and the mixing angle in the upper doublet by $\psi$. The
resulting complicated mixing matrix involving six mixing angles is greatly
simplified by application of the {\sc chooz} constraint. 
Then the flavour eigenstates are related to the four mass eigenstates in
vacuum through a unitary transformation,

\begin{equation}
\left[ \begin{array}{c} \nu_e \\ \nu_s \\ \nu_{\mu} \\ \nu_{\tau}
\end{array} \right] = U^v 
\left[ \begin{array}{c} \nu_1 \\ \nu_2 \\ \nu_3 \\ \nu_4 
\end{array} \right],
\end{equation}
where the superscript $v$ on the r.h.s. stands for vacuum.  Within the 
two doublet scheme the $4 \times 4$ unitary matrix, $U^v$, may be written as
\begin{equation}
U^v = \left( \begin{array}{cccc}
      (1-\epsilon^2)c_{\omega} & ~~~(1-\epsilon^2)s_{\omega} & \epsilon 
&\epsilon \\
     -(1-\epsilon^2)s_{\omega}-2\epsilon^2 c_{\omega} & 
(1-\epsilon^2)c_{\omega} -2\epsilon^2 s_{\omega}& \epsilon &\epsilon \\
\epsilon(s_{\omega}-c_{\omega})(c_{\psi}+s_{\psi})& 
-\epsilon(s_{\omega}+c_{\omega})(c_{\psi}+s_{\psi})& 
     (1-\epsilon^2)c_{\psi}-2\epsilon^2 s_{\psi} & 
     (1-\epsilon^2)s_{\psi}\\ 
\epsilon(s_{\omega}-c_{\omega})(c_{\psi}-s_{\psi})& 
-\epsilon(s_{\omega}+c_{\omega})(c_{\psi}-s_{\psi})& 
     -(1-\epsilon^2)s_{\psi}-2\epsilon^2 c_{\psi} & 
     (1-\epsilon^2)c_{\psi}\\ 

      \end{array} \right),
\end{equation}
where $s_{\omega} = \sin \omega$ and $c_{\omega} = \cos \omega$, etc. The
angles $\omega$ and $\psi$ can take values between $0$ and $\pi/2$.
The mixing matrix given above is unitary up to order $\epsilon^2$. Since 
$\epsilon$ is a small parameter, we do not need to go beyond this order.

\section{Three flavour MSW fits to Super-K data}
\begin{center}
S.R.Dugad, Mohan Narayan and Uma Shankar
\end{center}
There are several puzzles in the Super Kamiokande solar neutrino data:

(1) Higher energy bins ($12-14 MeV$) see less suppression than others.
The energy dependence of the observed neutrino flux does not any of the
three oscillation models- vacuum oscillations, small angle MSW or
large angle MSW.

(2) There is less suppression seen in the neutrino flux in the early
part of night than at midnight. This is at odds with the expectation 
that regeneration by the earth core is most effective at midnight.

This group intends to do MSW fits again using only the Gallium data
and the Super-K data on spectrum and time of night variations.

\section{$\nu$ decay solution to solar neutrino problem}
\begin{center}
S.Choubey, S.Goswami and D.Mazumdar
\end{center}

This group examines models where there is a mixing between $\nu_{e} 
\rightarrow \nu_{\mu}$ and in addition there is a decay of the heavier
neutrino $ \nu_{2L} \rightarrow \nu_{1R} + majoron$ . The survival
probability of $\nu_e$ is then given by (for $\Delta m^2 > 10^{-4} eV^2$)
\begin{equation}
P(\nu_e \rightarrow \nu_e) = (1- |U_{e2}|^2)^2 + |U_{e2}|^2 ~exp(-{L m_2
\over \tau_0 E})
\end{equation}

A fit with the superK spectrum data gives $\chi^2_{min}=14.87$ for
$15 d.o.f$ with the best fit values of the parameters
$|U_{e1}|^2 =0.63, m_2/tau_0 = 5.62\times 10^{-11} eV^2, \chi_n =12.6/5.15$
and is allowed at $46.08 \% $ C.L. w.r.t BP-98 standard solar model.

In comparison vacuum oscillations model gives $\chi^2_{min}=13.07$ for$15
d.o.f$ and MSW gives $\chi^2_{min}=17.23$ for$15 d.o.f$.

Detailed calculations of rates and spectral data analysis is planned.

\section{Role of collective neutrino forces in large scale structures}
\begin{center}
S.Mohanty, V.Sahni and A.M. Srivastava
\end{center}
Numerical simlations of cold+hot dark matter models 
show that  the dnesity perturbations do not match observations
at small scales ($100 kpc$) and large scales ($100 Mpc$). The CDM
component of dark matter clusters more than what is observed at scales
smaller than $100 kpc$ and the DM component free-streams and gives excess
power at scales larger than $100 Mpc$.

The proposal of this group is to study the collective neutrino
force on the CDM component which can be first order in $G_F$
\begin{equation}
F_c= G_F( - \nabla n_\nu + \vec v_c \times \nabla \times \vec J_\nu)
\end{equation}
This is different from 
the usual collisonal force which is $f= G_F^2 n_x$ which decouples
at very early times ($t\sim 1 sec$).

The Boltzmann equation of the CDM particles
\begin{equation}
\vec v_c \cdot \nabla_x f_c  + \vec F_c 
  \cdot \nabla_p f_c =0
\end{equation}
where the force term includes the collective weak interactions in
addition to gravity
\begin{equation}
F_c= - m_c\nabla_x \phi +
G_F( - \nabla n_\nu + \vec v_c \times \nabla \times \vec J_\nu)
\end{equation}

 A detailed analysis is proposed to check if the introduction of this
collective weak force can prevent excessive clustering of CDM.

\section{ Signature of Dirac vs Majorana nature of neutrinos
from HECR observations}
\begin{center}
K.R.S. Balaji, P. Bhattacharjee, S. Mohanty, S.N. Nayak
\end{center}

We examine the compatibilty
of the Wyler mechanism -which explains the observed UHECR above
the GZK cutoff energies ($10^19 GeV$) via annihilation of high energy neutrinos
with cosmic background neutrinos \cite{wyler} -
with the neutrino mass matrix suggested by the solar and atmospheric
neutrino problems.
If the neutrinos
are Dirac then there is a resonant spin flip by the magnetic fields of AGN's
from active to sterile chirality and the flux of UHECR would be decreased
by about $75 \%$.

{\it Resonant spin flip in AGN's}

Neutrinos are produced in AGNs or Radio Galaxies from pions arising from
collisions of protons after undergoing shock acceleration.
Each $\pi^+$ ultimately produces one each of $\nu_\mu$,  $\nu_\mu^c$ and
$\nu_e$. 
 The flavour
oscillations between different active species
will not change the number of real $Z'$s produced by annihilation
with the cosmic background neutrinos.  Oscillation from an active to
sterile neutrinos will which is possible in the case of Dirac neutrinos
will reduce the number of $Z$'s produced , and will have an observable
consequence in the decrease of UHECR produced $CB\nu$ annihilation.

The hamiltonian governing the propagation of two neutrino species
$(\nu_{L e}, N_{R,\alpha})$ where $\alpha = \mu $or $\tau   $  is given by

\be
i {d \over dt } \left( \barr{c} \nu_{L e } \\ N_{R, \alpha} \earr \right)
= \left(\barr{cc} V_c + V_n & \mu_{e \alpha} B\\ 
\mu_{e \alpha} B  & {\Delta m^2 \over 2 E} cos 2 \theta \earr \right)
\left( \barr{c} \nu_{L e } \\ N_{R, \alpha} \earr \right)
\label{h}
\ee

where $\Delta m^2 = (m_2^2 -m_1^2)$ and $\theta$ is the vaccumm mixing angle
between $\nu_e$ and $\nu_\alpha$. 
The matter potentials due to charged and neutral current are
 $V_c =2 \sqrt{2} G_F N_e$ and $V_n = - \sqrt{2} G_F N_n $  
respectively ( $N_e$ and $N_n$ are the number densities of electrons
and neutrons respectively).

The mixing angle in matter and magnetic
field is given by

\be
tan 2 \theta_M = {2 \mu_{e \alpha} B \over {\Delta m^2 cos 2\theta \over 2E}
 -(V_c +V_n)}
\label{tm}
\ee

The transition probability between $\nu_{L e} \rightarrow N_{R
\alpha}$ is given by

\be
P(\nu_{L e} \rightarrow N_{R\alpha})
= {1\over 2} -({1\over2} - e^{-\pi \kappa/2 } ) 
cos2 \theta cos 2 \theta_M(r_i)
\label{p}
\ee
where the matter mixing angle is evaluated at the point of production of
$\nu_{L e}$ and where the adiabaticity parameter $\kappa$ is given by
\be
\kappa(r_{res}) ={E \mu^2 B^2 \over \Delta m^2 cos 2\theta  |d ln(V_c
+V_n)/dr| }
\label{kappa}
\ee

We  now  calculate the amount of suppression of active neutrino flux due to
resonant spin-flavour precession (RSFP) in the AGN core model as a source
of the
high energy neutrinos.

 Active Galactic Nuclei are the most luminous objects
in the universe with luminosities in the range $10^{42} -10^{48} ergs/sec$.
These are supposed be powered by matter accreting into black holes
of masses $10^4 - 10^{10} M_{\odot} $.
 According to Szabo and Protheroe \cite{szabo} , the matter density
outside the core has the profile 
\bea
 \rho(r) &=& \rho_0 ~(r/R_s)^{-2.5} (1-0.1 (r/R_s)^{0.31}) ^{-1} ~~({10^{48}
ergs/sec \over L_{AGN}}) \nonumber\\
\rho_0 &=& 1.4\times 10^{-15} g/cm^3
\label{rho}
\eea
$R_s = 3\times 10^{11} (M_{AGN} / 10^8 M_{\odot})$ being the Schwarzchild
radius of the AGN. The magnetic field profile is given by
\bea
B(r) &=& B_0 ~(r/R_s)^{-1.75} (1- 0.1 (r/R_s)^{0.31})^{-0.5}
~~({10^{48}
ergs/sec \over L_{AGN}})^{1/2} \nonumber\\
B_0&=& 5.5 \times 10^{3} G
\label{B}
\eea

At the neutrino energies of interest, $E = 10^{20} eV$ ,
 the resonance condition
${\Delta m^2 cos 2\theta \over 2E} -(V_c +V_n)=0 $ 
is
attained in the AGN (at distances $r > 10 R_s$)
for $\Delta m^2 cos 2 \theta= 10^{-11} eV^2
( \rho/10^{-18} gm cm^{-3} )( E/ 10^{20} eV)  $. This mass difference
is in the same range as required for vaccumm oscillations 
(between $\nu_e$ and $\nu_\mu$) in order to solve the 
 solar neutrino 
neutrino problem.

The magnetic field at the core ($r =10 R_s$) is
$ B = 10 G$ , we have $ \mu B =1.7\times  10^{-26} eV$.
The mixing angle (\ref{tm}) at the core is  $cos\theta_M =
cos (\pi - arc tan ( 10^{-23}/ 10^{-24}) = -0.8 $ . 
The probability for conversion of $\nu_{L e}$ to a sterile $N_{R, \mu}$
is
\be
P(\nu_L \rightarrow N_{R \nu}) = 0.85   ;~~~~~~~~~~ 
\Delta m^2 = 5\times 10^{-11} eV^2 {\rm and}
sin^2 2\theta = 6 \times 10^{-3}.
\ee

Hence the flux of active $\nu_e$ from the AGN's at energies $10^{22} eV$
is reduced by $ (67 - 85) \%$ if the neutrinos are Dirac. If on the other
hand they are Majorana then the transitions will be between two active
species and there will be no reduction in the number of $Z$'s produced
when they anhilate the relic neutrinos.

For the case of $(\nu^c_{R, \mu})$ neutrinos produced at the core, 
the Hamiltonian governing its oscillation to a sterile $N_{L e}$ will be the
one in (\ref{h}) with extra minus signs on both the diagonal terms
and with $V_c=0$).
The resonance condition is  therefore the same and $\nu^c_{R, \mu}$
will convert to a sterile neutrino with a probabilty ($ 0.6- 0.8$).

For the  $\nu_{L,\mu}$ neutrinos produced at the AGN core there is
no matter resonanace. These will precess in the magnetic field
with a probability 
\be
P(\nu_{L \mu} \rightarrow N_R) = sin^2 (\mu B L) \rightarrow {1\over 2}
\ee

So the total flux of all three species of neutrinos produced at the core
reduces by about $75 \%$ if the neutrinos have a Dirac mass of about
$1 eV$.

\end{document}